\documentclass[nohyper,letterpaper,notoc]{JHEP}

\def\openone{\leavevmode\hbox{\small1\kern-3.8pt\normalsize1}}

\newcommand{\scr}{\scriptstyle}

\newcommand{\scri}{\scriptsize}

\def\bo{{\raise-.3ex\hbox{\large$\Box$}}}               
\def\face{{\raise.2ex\hbox{$\displaystyle \bigodot$}\mskip-2.2mu \llap {$\ddot
        \smile$}}}                                      
\def\tr{\mbox{\scri T}}                              

\def\Bar#1{\overline{#1}}                       
\def\ket#1{\left| #1\right\rangle}              
\def\leftrightarrowfill{$\mathsurround=0pt \mathord\leftarrow \mkern-6mu
        \cleaders\hbox{$\mkern-2mu \mathord- \mkern-2mu$}\hfill
        \mkern-6mu \mathord\rightarrow$}       
\def\dvec#1{\vbox{\ialign{##\crcr
        \leftrightarrowfill\crcr\noalign{\kern-1pt\nointerlineskip}
        $\hfil\displaystyle{#1}\hfil$\crcr}}}           

\def\beq{\begin{equation}}
\def\eeq{\end{equation}}

\def\beqx{\begin{displaymath}}
\def\eeqx{\end{displaymath}}

\def\beqa{\begin{eqnarray}}
\def\eeqa{\end{eqnarray}}
\def\NO{\nonumber}

\title{Three Family Type IIB Orientifold String Vacua with
Non-Abelian Wilson Lines}
\author{Mirjam Cveti\v c$^{a,c}$,  Michael Pl\"umacher$^a$, and
Jing Wang$^{a,b}$\\
$^a$ Department of Physics and Astronomy \\ 
University of Pennsylvania, Philadelphia PA 19104-6396, USA\\
$^b$Theory Division, Fermi National Accelerator Laboratory\\
P.O.Box 500, Batavia, IL 60510, USA\\
$^c$ Institute for Theoretical Physics, University of California\\
Santa Barbara, CA 93106, USA}
\abstract{
We address the implementation of non-Abelian Wilson lines in D=4 N=1
Type IIB orientifold constructions. We present an explicit
three-family example with the gauge group
(U(4)$\times$U(2)$\times$SU(2)$\times$SU(2))$^2\times
$(U(6)$\times$Sp(4))$^{2}$ and give the particle spectrum and the
trilinear superpotential. Emphasizing
the new subtleties associated with the introduction of non-Abelian Wilson
lines, we show that the Abelian gauge anomalies are cancelled by the
Green-Schwarz-type mechanism, and calculate the Fayet-Iliopoulos terms
and gauge coupling corrections. The analysis thus sets a stage for
further investigations of the phenomenological implications of this
model.
}
\preprint{UPR-865-T\\ NSF-ITP-99-132\\ 
FERMILAB-Pub-99/328-T\\ {\tt hep-th/9911021}}

\begin{document}

\section{Introduction}
Four-dimensional N=1 supersymmetric Type IIB
orientifolds\cite{ABPSS,berkooz,N1orientifolds,IbaneztypeI} provide a
domain of perturbative string vacua whose study of physics
implications is still at the early stages of
investigation~\cite{Ibanez,TyeKakush}.  In particular, there is the
need to further explore the existence of string vacua with
quasi-realistic features, i.e., those with gauge group close to that
of the standard model and massless spectra with three
families~\footnote{Recent explorations involve the construction of
Type IIB orientifolds with branes {\it and} anti-branes which break
supersymmetry, while keeping the models stable
(tachyon-free\cite{Sen,Antoniadisetal,UrangaRabadan}). A number of
quasi-realistic models were recently constructed in
\cite{Ibanezetal}.}.  In order to reduce the gauge group structure,
one mechanism involves the introduction of Wilson lines
\cite{IbaneztypeI,wilsonlinemodel}.
(Other mechanisms involve for example the blowing-up procedure \cite{CELW}
or the introduction of non-zero NS-NS two-form background fields
\cite{nonzeroBmodel}.)  

In particular, the introduction of Wilson lines that do not commute
with the discrete orbifold symmetry seems to be a promising mechanism
to reduce the gauge group structure and lead us a step closer to the
construction of quasi-realistic models~\cite{wilsonlinemodel}. We
address the consistent implementation of the symmetry actions of both
the discrete orientifold group and the Wilson line actions. In
addition, constraints arising from tadpole cancellation as well as the
requirement that the massless sector is free of non-Abelian anomalies,
severely restrict the allowed solutions.

We focus on a particular three family $Z_3\times Z_2\times Z_2$ Type
IIB orientifold model \cite{wilsonlinemodel} (a descendant of the
$Z_3$-orientifold \cite{ABPSS}, the first N=1 supersymmetric Type IIB
orientifold with three families and the gauge group
U(12)$\times$SO(8)). We address the possible Wilson lines, which
necessarily commute with the $Z_3$ but not all the $Z_2$ orbifold
actions.  We find that the consistency conditions (the absence of
tadpoles and non-Abelian gauge anomalies) allow for a very limited
number of possible models. We focus on the (only) example with gauge
group structure that encompasses that of the standard model and
present the massless spectrum and the trilinear
superpotential\footnote{The presented study builds on earlier
work \cite{wilsonlinemodel}. However, we find a different gauge group 
and massless spectrum.} (see
Section II).  In Section III, we explicitly demonstrate that the
Abelian gauge anomalies of the massless spectrum are cancelled by a
Green-Schwarz-type mechanism (proposed in \cite{U1anomaly} and
confirmed in \cite{dudas}); in particular, we explore the additional
subtleties that are associated with non-Abelian Wilson lines. In
Section IV we derive the form of the Fayet-Iliopoulos (FI) terms as
well as the corrections to the gauge kinetic functions due to the
blowing-up moduli.  In Section V we summarize results and point out
possible generalizations to address a general set-up for a consistent
construction of Type IIB orientifold solutions with non-Abelian Wilson
lines. In Appendix A we derive the massless matter spectrum and in
Appendix B the moduli of $Z_2$ twisted sectors.

\section{Orientifold constructions}
\subsection{$D=4$, $N=1$ orientifolds}
We compactify Type IIB theories on a six-torus $T^6$ and mod out a
discrete symmetry group $G_1$ and the world-sheet parity operation
$\Omega$, which could be accompanied by another discrete symmetry $G_2$,
i.e.\ the orientifold group is $G=G_1+\Omega G_2$.  Closure requires $\Omega g
\Omega g' \in G_1$ for $g,g'\in G_2$. In the following we always have
$G_1=G_2$.

The compactified tori are described by complex coordinates $X_i$,
$i=1,2,3$. The action of an orbifold group $Z_N$ on the compactified
dimensions can be summarized in a twist vector $v=(v_1,v_2,v_3)$,
\beq 
  g: X_i \to e^{2i\pi v_i}X_i\;,
\eeq
where the $v_i$'s are multiples of $1\over N$.  In our conventions,
${\cal N}=1$ supersymmetry requires that $v_1+v_2+v_3=1$.

Since the $\Omega$ projection relates left- and right-movers, it gives
rise to open strings \cite{horava}. Tadpole cancellation then requires
the inclusion of an even number of $D9$ branes and additional discrete
symmetries in the orientifold group may require the presence of
multiple sets of $D5$ branes for consistency.

An open string state can be written as $\ket{\Psi,ij}$ where $\Psi$ denotes the 
world-sheet state and $i,j$ the Chan-Paton states of the left and right end
points on a $D9$ or $D5$ brane. The elements $g\in G_1$ act on open string 
states as follows:
\beq
  g: \ket{\Psi,ij} \to 
  (\gamma_g)_{ii'}\ket{g\cdot\Psi,i'j'}(\gamma_g^{-1})_{j'j}\;.
  \label{ori1}
\eeq
Similarly, the elements of $\Omega G_1$ act as
\beq
  \Omega g: \ket{\Psi,ij} \to (\gamma_{\Omega g})_{ii'}\ket{\Omega g\cdot\Psi,j'i'}
  (\gamma_{\Omega g}^{-1})_{j'j}\;,
  \label{ori2}
\eeq
where we have defined $\gamma_{\Omega g}=\gamma_g\gamma_{\Omega}$, up to a phase, in
accordance with the usual rules for multiplication of group elements.
Note, that $\Omega$ exchanges the Chan-Paton indices. Hence, $(\Omega g)^2$
acts as
\beq
  (\Omega g)^2: \ket{\Psi,ij} \to [\gamma_{\Omega g}(\gamma_{\Omega g}^{-1})^{\tr}]_{ii'}
  \ket{(\Omega g)^2\cdot\Psi,i'j'}[\gamma_{\Omega g}^{\tr}\gamma_{\Omega g}^{-1}]_{j'j}\;.
\eeq

Since the $\gamma_g$ form a projective representation of the orientifold
group, consistency with group multiplication implies some conditions
on the $\gamma_g$. Consider the $G_1=G_2=Z_N$ case. If $g$ is the
generator of $Z_N$, we must have
\beq
 \gamma_g^N=\pm 1\;.
\eeq
Analogously, $\Omega^2=1$ implies that
\beq
  \gamma_{\Omega}=\pm \gamma_{\Omega}^{\tr}\;.
\eeq
Further, due to the closure relation $\Omega g \Omega g' \in G_1$ for $g,g'\in
G_2$, one must have
\beq
  (\gamma_g^k)^*=\pm\gamma_{\Omega}^*\gamma_g^k\gamma_{\Omega}\;.
  \label{gstar}
\eeq
It turns out that the tadpoles cancel, if we choose the plus sign for
$D9$ branes and the minus sign for $D5$ branes \cite{IbaneztypeI}.

In the $D9$ brane sector $\gamma_\Omega$ is symmetric and can be chosen real,
while it is antisymmetric in the $D5$ brane sectors and can be chosen
imaginary there. Hence, in a suitable basis
\beq
  \gamma_{\Omega,9}=\left(\begin{array}{cc}0&\openone_{16}\\ 
  \openone_{16}&0\end{array}\right)
  \qquad\mbox{and}\qquad
  \gamma_{\Omega,5}=\left(\begin{array}{cc}0&-i\openone_{16}\\ 
  i\openone_{16}&0\end{array}\right)\;,
  \label{omega}
\eeq
where the subscript $9$ or $5$ denotes the brane sector in which these
matrices are acting. Hence, in both the $D9$ and $D5$ brane sectors
eq.~(\ref{gstar}) yields
\beq
  (\gamma_g^k)^*=\gamma_{\Omega}\gamma_g^k\gamma_{\Omega}\;.
  \label{gstar2}
\eeq

Further, finiteness of string loop diagrams yields tadpole
cancellation conditions which constrain the traces of $\gamma_g$ matrices.
For example, in the case of $Z_3$ \cite{ABPSS}
\begin{equation}
\label{z3trace}
Tr(\gamma_{Z_3}) = -4\;. 
\end{equation}

Open string states, whose Chan-Paton matrices will be denoted by
$\lambda^{(i)}$, $i=0,\ldots,3$ in the following, give rise to space-time
gauge bosons ($i=0$) and matter states ($i=1,2,3$).  Gauge bosons in
the $D9$ brane sector arise from open strings beginning and ending on
$D9$ branes (the matter states will be discussed in Appendix
\ref{MATTER}).  Invariance of these states under the action of the
orientifold group requires
\beq
  \lambda^{(0)}=-\gamma_{\Omega,9}{\lambda^{(0)}}^{\tr}\gamma_{\Omega,9}^{-1} 
  \qquad\mbox{and}\qquad
  \lambda^{(0)}=\gamma_{g,9}\lambda^{(0)}\gamma_{g,9}^{-1}\;.
  \label{cond1}
\eeq
With eq.~(\ref{omega}) the first constraint implies that the
$\lambda^{(0)}$ are SO(32) generators, while the constraints from the
$\gamma_{g,9}$ will further reduce the group.

Similarly, in the $D5$ brane sectors the Chan-Paton matrices of the 
gauge bosons satisfy the following constraints:
\beq
  \lambda^{(0)}=-\gamma_{\Omega,5}{\lambda^{(0)}}^{\tr}\gamma_{\Omega,5}^{-1} 
  \qquad\mbox{and}\qquad
  \lambda^{(0)}=\gamma_{g,5_i}\lambda^{(0)}\gamma_{g,5_i}^{-1}\;.
\eeq
Due to the symplectic nature of $\gamma_{\Omega,5}$ in eq.~(\ref{omega}),
world-sheet parity yields Sp(32) generators for the gauge group,
which will be further reduced by the additional orientifold actions. 

Let us now consider the $Z_2\times Z_2\times Z_3$ orientifold, first
constructed in ref.~\cite{wilsonlinemodel}. Tadpole cancellation
implies that we have $32$ $D9$ branes and three different sets of $32$
$D5$ branes, which we will distinguish by an index $i$. The $D5_i$
brane fills the 4-dimensional space-time and the torus parametrized by
$X_i$. The $Z_3$ twist action on the tori is given by the twist vector
$v=({1\over 3},{1\over 3},{1\over 3})$ and, both in the $D9$ brane and
the $D5_i$ brane sectors, its action on Chan-Paton matrices is
generated by:
\beq
  \gamma_{Z_3}=\mbox{diag}\left(\omega\openone_{12},\openone_4,
           \omega^{-1}\openone_{12},\openone_4\right)\;,
  \quad\mbox{where}\quad\omega=\mbox{e}^{2\pi i/3}\;.
\eeq
This choice satisfies eqs.~(\ref{gstar2}) and (\ref{z3trace}).

The two $Z_2$'s act on the compactified coordinates $X_i$ ($i=1,2,3$)
in the following way:
\beqa
 R_1:&& X_1\to X_1\;,\quad X_{2,3}\to -X_{2,3}\quad\Rightarrow\quad
    v_{R_1}=\Big(0,{1\over2},{1\over2}\Big)\;, \label{R1def}\\[1ex]
 R_2:&& X_2\to X_2\;,\quad X_{1,3}\to -X_{1,3}\quad\Rightarrow\quad 
    v_{R_2}=\Big({1\over2},0,{1\over2}\Big)\;. \label{R2def}
\eeqa
The corresponding $\gamma$-matrices have to fulfill certain group
consistency conditions \cite{berkooz}:
\beqa
  \gamma_{R_i\Omega,s}^{\tr}&=&-{\cal C}_{i,s}\gamma_{R_i\Omega,s}\;,\label{cond2}\\
  \gamma_{R_i\Omega,s}{\gamma_{R_j\Omega,s}^{-1}}^{\tr}\gamma_{R_k\Omega,s}{\gamma_{\Omega,s}^{-1}}^{\tr}
  &=&{\cal C}_{0,s}{\cal C}_{3,s}\epsilon_{ijk}\;,\label{cond3}\\
  \gamma_{R_i\Omega,s}{\gamma_{\Omega,s}^{-1}}^{\tr}\gamma_{R_i\Omega,s}{\gamma_{\Omega,s}^{-1}}^{\tr}
  &=&{\cal C}_{0,s}{\cal C}_{i,s}\;,\label{cond4}\\
  \gamma_{R_i,s}\gamma_{R_j,s}\gamma_{R_k,s}&=&-{\cal C}_{0,s}{\cal C}_{3,s}\epsilon_{ijk}\;,
  \label{cond5}\\
  \gamma_{R_i,s}\gamma_{R_i,s}&=&{\cal C}_{0,s}{\cal C}_{i,s}\;,\label{cond6}
\eeqa
where $s=0$ and $s=1,2,3$ denote $D9$ branes and $D5_s$ branes,
respectively. The co-cycle ${\cal C}_{i,s}$ is equal to $-1$ for
$i=s$ and to $+1$ otherwise. 
Further, tadpole cancellation implies
\beq
  \mbox{Tr}\left(\gamma_{R_i,s}\right)=
  \mbox{Tr}\left(\gamma_{R_i,s}\gamma_{Z_3}\right)=0\;.
\eeq
One set of $\gamma_{R_i}$ matrices fulfilling
these constraints is given in table \ref{gamma_table}.

\TABLE{
\begin{tabular}{c|l}
     &$\gamma_{R_1,9}=\mbox{diag}\left(i\,\sigma_1\otimes\openone_6,
    i\,\sigma_1\otimes\openone_2,-i\,\sigma_1\otimes\openone_6,
    -i\,\sigma_1\otimes\openone_2\right) $\\[1ex]
  9  &$\gamma_{R_2,9}=\mbox{diag}\left(i\,\sigma_2\otimes\openone_6,
    i\,\sigma_2\otimes\openone_2,i\,\sigma_2\otimes\openone_6,
    i\,\sigma_2\otimes\openone_2\right) $\\[1ex]
     &$\gamma_{R_3,9}=\mbox{diag}\left(i\,\sigma_3\otimes\openone_6,
    i\,\sigma_3\otimes\openone_2,-i\,\sigma_3\otimes\openone_6,
    -i\,\sigma_3\otimes\openone_2 \right) $\\[1ex]
\hline
     &$\gamma_{R_1,5_1}=\mbox{diag}\left(i\,\sigma_1\otimes\openone_6,
    i\,\sigma_1\otimes\openone_2,-i\,\sigma_1\otimes\openone_6,
    -i\,\sigma_1\otimes\openone_2\right) $\\[1ex]
$\;5_1\;$&$\gamma_{R_2,5_1}=\mbox{diag}\left(\sigma_2\otimes\openone_6,
    \sigma_2\otimes\openone_2,-\sigma_2\otimes\openone_6,
    -\sigma_2\otimes\openone_2\right) $\\[1ex]
     &$\gamma_{R_3,5_1}=\mbox{diag}\left(\sigma_3\otimes\openone_6,
    \sigma_3\otimes\openone_2,\sigma_3\otimes\openone_6,
    \sigma_3\otimes\openone_2 \right) $\\[1ex]
\hline
     &$\gamma_{R_1,5_2}=\mbox{diag}\left(\sigma_1\otimes\openone_6,
    \sigma_1\otimes\openone_2,\sigma_1\otimes\openone_6,
    \sigma_1\otimes\openone_2\right) $\\[1ex]
$5_2$&$\gamma_{R_2,5_2}=\mbox{diag}\left(i\,\sigma_2\otimes\openone_6,
    i\,\sigma_2\otimes\openone_2,i\,\sigma_2\otimes\openone_6,
    i\,\sigma_2\otimes\openone_2\right) $\\[1ex]
     &$\gamma_{R_3,5_2}=\mbox{diag}\left(\sigma_3\otimes\openone_6,
    \sigma_3\otimes\openone_2,\sigma_3\otimes\openone_6,
    \sigma_3\otimes\openone_2 \right) $\\[1ex]
\hline
     &$\gamma_{R_1,5_3}=\mbox{diag}\left(\sigma_1\otimes\openone_6,
    \sigma_1\otimes\openone_2,\sigma_1\otimes\openone_6,
    \sigma_1\otimes\openone_2\right) $\\[1ex]
$5_3$&$\gamma_{R_2,5_3}=\mbox{diag}\left(-\sigma_2\otimes\openone_6,
    -\sigma_2\otimes\openone_2,\sigma_2\otimes\openone_6,
    \sigma_2\otimes\openone_2\right) $\\[1ex]
     &$\gamma_{R_3,5_3}=\mbox{diag}\left(i\,\sigma_3\otimes\openone_6,
    i\,\sigma_3\otimes\openone_2,-i\,\sigma_3\otimes\openone_6,
    -i\,\sigma_3\otimes\openone_2 \right) $
\end{tabular}
\caption{Representation of the $Z_2\times Z_2$ twist matrices in 
   the $D9$ and $D5_i$ brane sectors.\protect\label{gamma_table}}
}

Consider now the Chan-Paton matrix for the gauge fields in the $D9$ brane
sector. Inserting $\gamma_{Z_3,9}$ and the $\gamma_{R_i,9}$ in eq.~(\ref{cond1})
yields the following form for $\lambda^{(0)}$
\beq
\lambda^{(0)}=\left(\begin{array}{cccc}
\openone_2\otimes B_1 &  0  &  0  & 0 \\[1ex]
   0  & \openone_2\otimes B_2 & 0  & i\sigma_2\otimes S_1 \\[1ex]
   0  &  0  &-\openone_2\otimes B_1^{\tr}&  0  \\[1ex]
   0  &  -i\sigma_2\otimes S_2  & 0  & -\openone_2\otimes B_2^{\tr}
  \end{array}\right)\;.\label{lambda0}
\eeq
Here, $B_1$ is a general $6\times6$ matrix, corresponding to the
adjoint representation of U$(6)$. $B_2$ is a general $2\times2$
matrix and $S_1$ and $S_2$ are symmetric $2\times2$ matrices, i.e.\
$B_2$, $S_1$ and $S_2$ form an adjoint representation of Sp(4).

In the $D5_i$ brane sectors, the Chan-Paton matrices for the gauge
fields have a similar structure, i.e.\ we get three additional copies
of the gauge group U(6)$\times$Sp(4).

\TABLE[b]{
\begin{tabular}{c||c||c|c}
&$\mbox{Tr}\left(\gamma_g\lambda^{(0)}\right)$&
\multicolumn{2}{c}{$\mbox{Tr}\left(\left(\gamma_g\right)^{-1}
\left(\lambda^{(0)}\right)^2\right)$}\\[1ex]
\hline
 $\gamma_g$ & U(6) & U(6) & Sp(4)\\[1ex]
\hline
 & & &\\[-2ex]
$\gamma_{Z_3}$ &
 $2i\sqrt{3}\mbox{Tr}B_1$ & $-2\mbox{Tr}B_1^2$ & 
 $4\mbox{Tr}\left(B_2^2+S_2S_1\right)$\\[1ex]
$\gamma_{Z_3}^2$ &
 $-2i\sqrt{3}\mbox{Tr}B_1$ & $-2\mbox{Tr}B_1^2$ & 
 $4\mbox{Tr}\left(B_2^2+S_2S_1\right)$
\end{tabular}
\caption{Contributions of the different gauge groups to the traces 
  relevant for anomaly cancellation, FI-terms and gauge coupling
  corrections. $B_1$ represents a U$(6)$ generator, while $B_2$
  and the $S_i$ generate the Sp$(4)$ (See the detailed explanation 
  after eq.~(\ref{lambda0})). All the other traces, e.g.\ those
  involving $Z_2$ twists $\gamma_{R_i}$, vanish.
  \protect\label{traces_table}}
}

\subsection{Wilson lines}
In order to further break the gauge symmetry we introduce discrete
Wilson lines, which can be written as a matrix $\gamma_W$ acting on the
Chan-Paton matrices. A Wilson line along the two-torus $X_i$ has to
satisfy the following algebraic consistency conditions
\beq
  \left(\gamma_{Z_3}\gamma_W\right)^3=+1\;,\qquad
  \left(\gamma_{R_j,9}\gamma_W\right)^2=-1\;,\qquad
  \left(\gamma_{R_j,5_i}\gamma_W\right)^2=+1\qquad\mbox{for }j\ne i\;.
  \label{Wcond1}
\eeq
These conditions and eqs.~(\ref{cond5}) and (\ref{cond6}) then imply
that $\gamma_W$ and $\gamma_{R_i}$ have to commute in the $D9$ and $D5_i$ brane
sectors,
\beq
  \left[\gamma_{R_i,9},\gamma_W\right]=0\;,\qquad
  \left[\gamma_{R_i,5_i},\gamma_W\right]=0\;.
\eeq
This is due to the fact that $\gamma_{R_i}$ does not act on the
compactified coordinate $X_i$ (cf.~eqs.~(\ref{R1def}) and
(\ref{R2def})). Further, tadpole cancellation requires
\beqa
  &&\mbox{Tr}\left(\gamma_{Z_3}\right)=\mbox{Tr}\left(\gamma_{Z_3}\gamma_W\right)
  =\mbox{Tr}\left(\gamma_{Z_3}\gamma_W^2\right)=-4\;,\label{Wcond2}\\[1ex]
  &&\mbox{Tr}\left(\gamma_{Z_3}\gamma_W\gamma_{R_j,s}\right)=
  \mbox{Tr}\left(\gamma_{Z_3}\gamma_W^2\gamma_{R_j,s}\right)=0\;,\quad
  \mbox{for }s=9\mbox{ or }5_i\;.\label{Wcond3}
\eeqa

In the following we will restrict ourselves to the case of one Wilson
line along the third two-torus. It turns out that 
eqs.~(\ref{Wcond1})--(\ref{Wcond3}) allow only one Wilson line (up 
to equivalent representations) which yields a gauge group containing
the standard model as a subgroup:
\beqa
  &&\gamma_W=\mbox{diag}\left(W_1,W_2,W_1^{-1},W_2^{-1},W_3,W_3^{-1},
  W_1^{-1},W_2^{-1},W_1,W_2,W_3^{-1},W_3\right)\;,\\[1ex]
  &&\mbox{where }W_1=\mbox{diag}\left(\omega,\omega,1\right)\;,\qquad
  W_2=\openone_3\;,\qquad W_3=\mbox{diag}\left(\omega,1\right)\;.
\eeqa
Since we place all the $D5$ branes at the origin, this Wilson line
acts only in the $D9$ and $D5_3$ brane sectors.  The gauge bosons in
these two sectors have to obey one additional constraint,
\beq
  \lambda^{(0)}=\gamma_W\lambda^{(0)}\gamma_W^{-1}\;.
\eeq
The generator $B_1$ of the U(6) in eq.~(\ref{lambda0}) therefore gets
reduced to
\beq
  B_1=\left(\begin{array}{cc}
     \tilde{B}_1 & 0\\[1ex]
         0 & \tilde{B}_2
   \end{array}\right)\;,
\eeq
where $\tilde{B}_1$ and $\tilde{B}_2$ are general $2\times 2$ and
$4\times4$ matrices, respectively. Hence, the U(6) gets broken down
to a U(4)$\times$U(2).  Similarly, the Sp(4) generators $B_2$ and
$S_i$ are reduced to
\beq
  B_2=\left(\begin{array}{cc}
         W_1 & 0   \\[1ex]
         0   & Z_1 
   \end{array}\right)\;, \qquad
  S_1=\left(\begin{array}{cc}
         W_2 & 0   \\[1ex]
         0 & Z_2 
   \end{array}\right)\;, \qquad
  S_2=\left(\begin{array}{cc}
         W_3 & 0   \\[1ex]
         0 & Z_3 
   \end{array}\right)\;.
\eeq
Here, the $W_i$ and $Z_i$ generate two SU(2)'s.  

\TABLE{
\begin{tabular}{c|c|c|c}
Sector & Gauge group & Field & Representation \\[1ex]
\hline
99 & U(4)$\times$U(2)$\times$SU(2)$\times$SU(2) 
   & $\chi_k^{(0)}$         & $3\times(6,1,1,1)(+2,0)$\\[0.5ex]
 & & $\psi_k^{(0)}$       & $3\times(\bar{4},1,1,2)(-1,0)$\\[0.5ex]
 & & $\eta^{(0)}$         & $(1,1,1,1)(0,+2)$\\[0.5ex]
 & & $\tilde{\eta}^{(0)}$ & $(1,2,2,1)(0,-1)$\\[1ex]
\hline
$5_i5_i$ & U(6)$\times$Sp(4) 
          & $\chi_k^{(i)}$   & $3\times(15,1)(+2)$\\[0.5ex]
$i=1,2$ & & $\psi_k^{(i)}$ & $3\times(\bar{6},4)(-1)$\\[1ex]
\hline
$5_35_3$ & U(4)$\times$U(2)$\times$SU(2)$\times$SU(2) 
   & $\chi_k^{(3)}$   & $3\times(6,1,1,1)(+2,0)$\\[0.5ex]
 & & $\psi_k^{(3)}$ & $3\times(\bar{4},1,1,2)(-1,0)$\\[0.5ex]
 & & $\eta^{(3)}$   & $(1,1,1,1)(0,+2)$\\[0.5ex]
 & & $\tilde{\eta}^{(3)}$ & $(1,2,2,1)(0,-1)$\\[1ex]
\hline
$95_i$ & $\left[\mbox{U(4)}\times \mbox{U(2)}\times \mbox{SU(2)}
          \times \mbox{SU(2)}\right]_9$
          & $P^{(0\,i)}$ & $(1,2,1,1;6,1)(0,+1;+1)$\\[0.5ex]
$i=1,2$ & $\times\left[\mbox{U(6)}\times \mbox{Sp(4)} \right]_{5_i}$  
          & $Q^{(0\,i)}$ & $(4,1,1,1;6,1)(+1,0;+1)$\\[0.5ex]
 &        & $R^{(0\,i)}$ & $(1,2,1,1;1,4)(0,-1;0)$\\[0.5ex]
 &        & $S^{(0\,i)}$ & $(\bar{4},1,1,1;1,4)(-1,0;0)$\\[0.5ex]
 &        & $T^{(0\,i)}$ & $(1,1,2,1;\bar{6},1)(0,0;-1)$\\[0.5ex]
 &        & $U^{(0\,i)}$ & $(1,1,1,2;\bar{6},1)(0,0;-1)$\\[1ex]
\hline
$95_3$ & $\left[\mbox{U(4)}\times\mbox{U(2)}\times\mbox{SU(2)}\times
       \mbox{SU(2)}\right]_9$
   & $Q^{(0\,3)}$ & $(4,1,1,1;4,1,1,1)(+1,0;+1,0)$\\[0.5ex]
 & $\qquad\times\left[\mbox{U(4)}\times\mbox{U(2)}\times\mbox{SU(2)}
  \times\mbox{SU(2)}\right]_{5_3}$ 
   & $S^{(0\,3)}$ & $(\bar{4},1,1,1;1,1,1,2)(-1,0;0,0)$\\[0.5ex]
 & & $U^{(0\,3)}$ & $(1,1,1,2;\bar{4},1,1,1)(0,0;-1,0)$\\[1ex]
\hline
$5_15_2$ & $\left[\mbox{U(6)}\times\mbox{Sp(4)} \right]_{5_1}
      \times\left[\mbox{U(6)}\times\mbox{Sp(4)} \right]_{5_2}$ 
   & $Q^{(1\,2)}$ & $(6,1;6,1)(+1;+1)$\\[0.5ex]
 & & $S^{(1\,2)}$ & $(\bar{6},1;1,4)(-1;0)$\\[0.5ex]
 & & $U^{(1\,2)}$ & $(1,4;\bar{6},1)(0;-1)$\\[1ex]
\hline
$5_35_i$ & $\left[\mbox{U(4)}\times\mbox{U(2)}\times\mbox{SU(2)}
   \times\mbox{SU(2)}\right]_{5_3}$
          & $P^{(3\,i)}$ & $(1,2,1,1;6,1)(0,+1;+1)$\\[0.5ex]
$i=1,2$ & $\times\left[\mbox{U(6)}\times\mbox{Sp(4)}\right]_{5_i}$
          & $Q^{(3\,i)}$ & $(4,1,1,1;6,1)(+1,0;+1)$\\[0.5ex]
        & & $R^{(3\,i)}$ & $(1,2,1,1;1,4)(0,-1;0)$\\[0.5ex]
        & & $S^{(3\,i)}$ & $(\bar{4},1,1,1;1,4)(-1,0;0)$\\[0.5ex]
        & & $T^{(3\,i)}$ & $(1,1,2,1;\bar{6},1)(0,0;-1)$\\[0.5ex]
        & & $U^{(3\,i)}$ & $(1,1,1,2;\bar{6},1)(0,0;-1)$\\[1ex]
\end{tabular}
\caption{The massless spectrum from the open string states. The
indices $k$ on the $\chi^{(s)}$ and $\psi^{(s)}$ fields are family
indices, i.e.\ $k=1,2,3$.
  \protect\label{matter_table}}
}
\clearpage

In the $5_1$ and $5_2$ sectors the gauge groups remain unbroken. 
The action of Wilson lines on the matter states is discussed in
Appendix \ref{MATTER} and the resulting spectrum of the $Z_3
\times Z_2 \times Z_2$ model with one Wilson line is given in table
\ref{matter_table}. 

The tree-level Yukawa superpotential of the massless states
reads\footnote{The nonzero terms in the superpotential are determined
by gauge invariance and standard techniques of superconformal field
theory in the closed superstring sector \cite{superp}.}
\beqa
{\cal W}&=& \sum\limits_{s=0,1,2,3}\;\sum\limits_{i\ne j\ne k}
       \chi_i^{(s)}\psi_j^{(s)}\psi_k^{(s)}
 + \chi_i^{(0)}S^{(0i)}S^{(0i)} 
 + \varepsilon_{ijk}\chi_i^{(j)}S^{(jk)}S^{(jk)}
 + \psi_i^{(0)}Q^{(0i)}U^{(0i)}\NO\\[1ex]
&& + \varepsilon_{ijk}\psi_i^{(j)}Q^{(jk)}U^{(jk)}
+ \varepsilon_{ij3}\chi_i^{(j)}U^{(3j)}U^{(3j)}
+ \varepsilon_{ij3}\chi_i^{(j)}T^{(3j)}T^{(3j)}
+ \varepsilon_{ij3}\psi_i^{(j)}P^{(3j)}R^{(3j)}\NO\\[1ex]
&& + \varepsilon_{ij3}\psi_i^{(j)}Q^{(3j)}S^{(3j)}
 + \chi_i^{(i)}U^{(0i)}U^{(0i)}
 + \chi_i^{(i)}T^{(0i)}T^{(0i)}
 + \psi_i^{(i)}P^{(0i)}R^{(0i)} 
 + \psi_i^{(i)}Q^{(0i)}S^{(0i)} \NO\\[1ex]
&& + \sum\limits_{s=0,3}\left(
   P^{(si)}R^{(sj)}S^{(ij)} + Q^{(si)}S^{(sj)}S^{(ij)}
   + T^{(si)}T^{(sj)}Q^{(ij)} + U^{(si)}U^{(sj)}Q^{(ij)}\right)\;,
\eeqa
where summation over repeated indices is understood and we defined
$S^{(ij)}\equiv U^{(ji)}$. Further, we have suppressed the actual
Yukawa couplings and determined only trilinear couplings. Note, that
$\eta^{(s)}$ and $\tilde{\eta}^{(s)}$, the extra matter states from
$\lambda^{(3)}$ in the $9$ and $5_3$ sectors (cf.~Appendix A), do not have
trilinear couplings.

\TABLE{
\begin{tabular}{c||c|c||c|c|c|c}
&\multicolumn{2}{c}{$\mbox{Tr}\left(\gamma_g\lambda^{(0)}\right)$}&
\multicolumn{4}{c}{$\mbox{Tr}\left(\left(\gamma_g\right)^{-1}
\left(\lambda^{(0)}\right)^2\right)$}\\[1ex]
\hline
 $\gamma_g$ & U(4) & U(2) & U(4) & U(2) & SU(2) & SU(2)\\[1ex]
\hline
 & & & & & &\\[-2ex]
$\scr\gamma_{Z_3}$ &
 $\scr 2i\sqrt{3}\mbox{\scri Tr}\tilde{B}_2$ & 
 $\scr 2i\sqrt{3}\mbox{\scri Tr}\tilde{B}_1$ &
 $\scr -2\mbox{\scri Tr}\tilde{B}_2^2$ &
 $\scr -2\mbox{\scri Tr}\tilde{B}_1^2$ & 
 $\scr 4\left(W_1^2+W_2W_3\right)$ &
 $\scr 4 \left(Z_1^2+Z_2Z_3\right)$\\[1ex]
$\scr \gamma_{Z_3}^2$ &
 $\scr -2i\sqrt{3}\mbox{\scri Tr}\tilde{B}_2$ & 
 $\scr -2i\sqrt{3}\mbox{\scri Tr}\tilde{B}_1$ &
 $\scr -2\mbox{\scri Tr}\tilde{B}_2^2$ &
 $\scr -2\mbox{\scri Tr}\tilde{B}_1^2$ & 
 $\scr 4\left(W_1^2+W_2W_3\right)$ &
 $\scr 4\left(Z_1^2+Z_2Z_3\right)$\\[1ex]
\hline
 & & & & & &\\[-2ex]
$\scr \gamma_{Z_3}\gamma_W$ &
 $\scr 2i\sqrt{3}\mbox{\scri Tr}\tilde{B}_2$ &
 $\scr -i\sqrt{3}\mbox{\scri Tr}\tilde{B}_1$ &
 $\scr -2\mbox{\scri Tr}\tilde{B}_2^2$ &
 $\scr \mbox{\scri Tr}\tilde{B}_1^2$ & 
 $\scr -2\left(W_1^2+W_2W_3\right)$ & 
 $\scr 4\left(Z_1^2+Z_2Z_3\right)$\\[1ex]
$\scr \gamma_{Z_3}^2\gamma_W$ &
 $\scr -2i\sqrt{3}\mbox{\scri Tr}\tilde{B}_2$ &
 $\scr i\sqrt{3}\mbox{\scri Tr}\tilde{B}_1$ &
 $\scr -2\mbox{\scri Tr}\tilde{B}_2^2$ &
 $\scr \mbox{\scri Tr}\tilde{B}_1^2$ & 
 $\scr -2\left(W_1^2+W_2W_3\right)$  & 
 $\scr 4\left(Z_1^2+Z_2Z_3\right)$\\[1ex]
$\scr \gamma_{R_3}\gamma_W$ &
 {\scri 0} & 
 {\scri 0} & 
 {\scri 0} & 
 $\scr -2\sqrt{3}\mbox{\scri Tr}\tilde{B}_1^2$ &
 $\scr -2\sqrt{3}\left(W_1^2+W_2W_3\right)$ & 
 {\scri 0} \\[1ex]
$\scr \gamma_{Z_3}\gamma_{R_3}\gamma_W$ &
 {\scri 0} & 
 $\scr -3i\mbox{\scri Tr}\tilde{B}_1$ & 
 {\scri 0} &
 $\scr \sqrt{3}\mbox{\scri Tr}\tilde{B}_1^2$ & 
 $\scr -2\sqrt{3}\left(W_1^2+W_2W_3\right)$ & 
 {\scri 0} \\[1ex]
$\scr \left(\gamma_{Z_3}\gamma_{R_3}\right)^5\gamma_W$ &
 {\scri 0} & 
 $\scr 3i\mbox{\scri Tr}\tilde{B}_1$ & 
 {\scri 0} &
 $\scr \sqrt{3}\mbox{\scri Tr}\tilde{B}_1^2$ &  
 $\scr -2\sqrt{3}\left(W_1^2+W_2W_3\right)$ & 
 {\scri 0} \\[1ex]
\hline
 & & & & & &\\[-2ex]
$\scr \gamma_{Z_3}\gamma_W^2$ &
 $\scr 2i\sqrt{3}\mbox{\scri Tr}\tilde{B}_2$ &
 $\scr -i\sqrt{3}\mbox{\scri Tr}\tilde{B}_1$ &
 $\scr -2\mbox{\scri Tr}\tilde{B}_2^2$ &
 $\scr \mbox{\scri Tr}\tilde{B}_1^2$ & 
 $\scr -2\left(W_1^2+W_2W_3\right)$ & 
 $\scr 4\left(Z_1^2+Z_2Z_3\right)$\\[1ex]
$\scr \gamma_{Z_3}^2\gamma_W^2$ &
 $\scr -2i\sqrt{3}\mbox{\scri Tr}\tilde{B}_2$ &
 $\scr i\sqrt{3}\mbox{\scri Tr}\tilde{B}_1$ &
 $\scr -2\mbox{\scri Tr}\tilde{B}_2^2$ &
 $\scr \mbox{\scri Tr}\tilde{B}_1^2$ & 
 $\scr -2\left(W_1^2+W_2W_3\right)$ & 
 $\scr 4\left(Z_1^2+Z_2Z_3\right)$\\[1ex]
$\scr \gamma_{R_3}\gamma_W^2$ &
 {\scri 0} & 
 {\scri 0} & 
 {\scri 0} & 
 $\scr 2\sqrt{3}\mbox{\scri Tr}\tilde{B}_1^2$ & 
 $\scr 2\sqrt{3}\left(W_1^2+W_2W_3\right)$ & 
 {\scri 0} \\[1ex]
$\scr \gamma_{Z_3}\gamma_{R_3}\gamma_W^2$ &
 {\scri 0} & 
 $\scr 3i\mbox{\scri Tr}\tilde{B}_1$ & 
 {\scri 0} &
 $\scr -\sqrt{3}\mbox{\scri Tr}\tilde{B}_1^2$ &  
 $\scr 2\sqrt{3}\left(W_1^2+W_2W_3\right)$ & 
 {\scri 0} \\[1ex]
$\scr \left(\gamma_{Z_3}\gamma_{R_3}\right)^5\gamma_W^2$ &
 {\scri 0} & 
 $\scr -3i\mbox{\scri Tr}\tilde{B}_1$ & 
 {\scri 0} &
 $\scr -\sqrt{3}\mbox{\scri Tr}\tilde{B}_1^2$ &  
 $\scr 2\sqrt{3}\left(W_1^2+W_2W_3\right)$ & 
 {\scri 0}
\end{tabular}
\caption{Contributions of the different gauge groups in the $D9$ and 
  $D5_3$ brane sectors to the traces relevant for anomaly cancellation,
  FI-terms and gauge coupling corrections for the model with the
  Wilson line. $\tilde{B}_1$ and $\tilde{B}_2$ are the U(2) and
  U(4) generators and the $W_i$ and $Z_i$ form the SU(2)
  generators. In the $D5_{1,2}$ brane sectors the traces remain
  unchanged, i.e.\ they can be found in table \ref{traces_table}.
  \protect\label{traces_w_table}}
}

\section{Chern-Simons terms and Anomaly Cancellation}

It is obvious from the spectrum that the U$(1)$ factors in various
sectors are anomalous at the effective field theory level.  Therefore,
it is important to check that the Abelian anomalies cancel through the
generalized Green-Schwarz mechanism \cite{U1anomaly}.
 
Before we demonstrate the cancellation of the U(1) anomalies
explicitly, we turn to a discussion of the NS-NS sector moduli fields
of the $Z_2\times Z_2\times Z_3$ orientifold.  There are three
untwisted sector moduli associated with the scaling deformation of
each two-torus. If we denote the complexified bosonic coordinate
associated with the $J^{th}$ two-torus as $X^J$, the three moduli
fields are represented by $\partial_zX^J\partial_{\bar z} \bar{X}^J$
($J=1,2,3$) vertices.

The twisted sector moduli fields arise from different twisted sectors
and are associated with the fixed points of the $Z_2\times Z_2\times
Z_3$ orientifold.  The twisted sectors are generated by the
multiplications of elements of $Z_3$ ($\{ 1,~ 
\theta \equiv \mbox{diag}(e^{{2\pi i}\over 3},e^{{2\pi i}\over 3}, e^{{2\pi
i}\over 3}), ~\theta ^2 \}$), and the two $Z_2$'s, generated by
$R_1=\mbox{diag}(1,-1,-1)$ and $R_2=\mbox{diag}(-1,1,-1)$,
respectively.  Each sector is specified by a certain number of fixed
points (and/or fixed two-tori). However, not all the fixed points in
these sectors are in one-to-one correspondence with the 
physical states, such as blowing-up moduli. Generally, in each sector
only particular combinations of blowing up-modes that are invariant
under the action of the remaining discrete rotations survive.

The blowing-up modes of the $Z_3$ twisted sectors can be identified in
the following way. The $\theta$ sector has $27=3\times 3\times 3$
fixed points, and twisted fields $\sigma_i^J$ are associated with
the $i^{th}$ fixed point ($i=1,2,3$) of the $J^{th}$ two-torus $T^2$
($J=1,2,3$).  The $Z_3$ orientifold would have 27 blowing-up moduli
fields of the form $\phi_{i,j,k}\equiv \sigma_i^1 \sigma_j^2
\sigma_k^3$, where $(i,j,k)=\{1,2,3\}$ run over the three fixed points
on each torus.  However, the invariance of the states under the two
discrete $Z_2$-rotations $R_1$ and $R_2$ implies that only nine
combinations of the original blowing-up modes are physical. Those are
(a) the blowing-up mode at the origin, (b) three blowing-up modes that are
symmetric combinations of $\phi_{i,j,k}$ with one of the indices being
$2$ and the other two indices fixed to be $1$ and its mirror image
under $Z_2$ twists, (c) three blowing-up modes with a symmetric
combinations of $\phi_{i,j,k}$ with two of the indices being $2$ and
the other index fixed to be $1$ and its three mirror images under
$Z_2$ twists and (d) two blowing-up modes which are symmetric and
antisymmetric combinations of $\phi_{i,j,k}$ with $i=j=k=2$ and its
five images under $Z_2$ twists.

The $16$ blowing-up modes of each of the $R_i$ twists are combined to
generate $6$ physical moduli that are also invariant under the $Z_3$
action (cf.\ Appendix B).  The order $6$ twists $\theta R_i$
$(i=1,2,3)$ each have two blowing-up moduli ($R_3\equiv R_1
R_2=(-1,-1,0)$), which can be seen as follows, using $\theta R_{3}$ as
an example.  The $\theta R_3$ rotation is generated by
$\mbox{diag}(e^{\frac{-2 \pi i}{6}}, e^{\frac{-2 \pi i}{6}},
e^{\frac{2\pi i}{3}})$. The order-$6$ twists in both the first and
second tori generate one fixed point (at the origin in each of the
two-tori). The third torus has three fixed points due to the
order-three twist. However, the $R_1$ and $R_2$ invariance selects out
the two physical modes, $\rho_{1,1,1}$ and $\rho_{1,1,\{2,3\}} \equiv
\frac{1}{\sqrt{2}}(\rho_{1,1,2} + \rho_{1,1,3})$, which combines the
two states that are mirror pairs under $R_1$ and $R_2$.

\subsection{Chern-Simons terms}
Anomaly cancellation is ensured by the existence of the Chern-Simons (CS)
term of the effective theory \cite{douglasmoore}, which is of the form 
\begin{equation}
I_{CS} \sim \sum_k \int d^4 x ~ C_k \wedge e^{F}\;,
\end{equation}
where $C_{k}$ are the 2-form Ramond-Ramond moduli fields arising from
the $k$th twisted sector and $F$ is the field strength of the gauge 
bosons. The first order expansion in $F$ gives the term 
\begin{equation}
\label{1thCS}
\sum_{k} \int d^4x ~C_{k} \wedge \mbox{Tr}(\gamma_k \lambda^{(0)}) F\;,
\end{equation}
where the coefficient $\mbox{Tr}(\gamma_k \lambda^{(0)})$ is associated to 
the orbifold action and the Chan-Paton matrix of the gauge bosons. 
The second order expansion gives the coupling between the twisted
sector RR fields and $F \tilde{F}$, with the prefactor 
$\mbox{Tr}((\gamma_k)^{-1} {\lambda^{(0)}}^2)$. These two couplings are
responsible for the cancellation of field theoretical U(1) 
anomalies \cite{U1anomaly}. 

The particular combinations of RR twisted sector fields that are
responsible for U$(1)$ anomaly cancellations 
require careful study. For example, the explicit trace calculation
for the $Z_2 \times Z_2 \times Z_3$ orientifold without Wilson line
(table \ref{traces_table}) reveals that $\mbox{Tr}(\gamma_k \lambda^{(0)})$
vanishes for all the twists except those of $Z_3$. It implies that the
RR twisted moduli that are responsible for anomaly cancellation
are from the $Z_3$ twisted sectors only. For the U$(1)$'s arising from
the $D9$ branes, which extend over all six compactified dimensions,
all the physical RR field are involved. Namely, the RR field for
anomaly cancellation in the $D9$ brane sector $B^9$ takes the form
\begin{equation}
B^{9} \propto \sum_{i,j,k} \psi_{i,j,k} \;,
\end{equation}
where $\psi_{i,j,k}$ are the RR partner of the twisted sector moduli
$\phi_{i,j,k}$ of $Z_3$. For the $D5_3$ branes sitting at the
origin, extending over the $i$th complex coordinate, the relevant RR 
field for anomaly cancellation $B^{5_3}$ takes the form  
\begin{equation}
B^{5_3} \propto \sum_{i} ~\psi_{1,1,i}\;.
\end{equation}
$B_{5_1}$ and $B_{5_2}$ have similar forms. The situation in the
model with Wilson line is more complicated and will be discussed later. 

\subsection{Anomaly Cancellation}
The U$(1)$ mixed anomalies in the effective theory are cancelled by
the exchange of twisted sector RR fields between the gauge bosons, as
discussed in \cite{U1anomaly}. The amplitude for this process is given
by the expression
\begin{equation} 
\label{Alm}
A_{lm}^{\alpha \beta} = \frac{i}{|P|} \sum_{k} ~C_{k}^{\alpha
\beta}(v) \mbox{Tr}(\gamma_{k}^{\alpha} \lambda^{(0)}_{l})
\mbox{Tr}((\gamma_{k}^{\beta})^{-1} (\lambda^{(0)}_{m})^2) \;, 
\end{equation} 
where $\alpha,~\beta=9,~5_i$ denote the brane sectors from which the U$(1)$ group and the non-Abelian group arise; $P$ is the order of the orbifold group, 
which for the $Z_2 \times Z_2 \times Z_3$ orientifold takes the value 
$2\times (2 \times 2 \times 3)= 24$, and $k$ runs over all the twisted
sectors. The factor $C_{k}^{\alpha\beta}$ arises from string tadpole 
calculations. In the case of $\alpha = \beta$, 
\begin{equation} 
C_{k}^{\alpha \beta}(v) = (-1)^k \Pi_{a=1}^{3} ~2 \sin(\pi k v_a)\;. 
\end{equation} 
$Z_3$ action gives $C_{1}^{\alpha \beta}(v) = - C_{2}^{\alpha \beta}(v)=
-3\sqrt{3}$. If $\alpha=9$ and $\beta=5_i$, 
\begin{equation} 
C_{k}^{9 5_i}(v) = (-1)^k 2 \sin(\pi k v_i)\;, 
\end{equation} 
which gives $ C_{1}^{9 5_i}(v)= - C_{2}^{9 5_i}(v) = -\sqrt{3}$ for
$Z_3$. If $\alpha=5_i$ and $\beta=5_j$, 
\beq
  C_{k}^{5_i 5_j}(v)= (-1)^k 2 \sin(\pi k v_a)\;,\quad
  \mbox{where } a \neq i \neq j\;.
\eeq
Hence, the order $3$ twists of the $Z_2 \times Z_2 \times Z_3$
orientifold give $C_{1}^{5_i 5_j}(v)= - C_{2}^{5_i 5_j}(v) =
-\sqrt{3}$.

When a background Wilson line is introduced into the world-volume of
the $D9$ branes, the fixed points of the orbifold action generally
split into different sets, each of which feels different gauge
monodromy. The orbifold action on the Chan-Paton matrices
is modified accordingly at each fixed point, and the total amplitude
$A_{lm}$ of the RR twisted moduli exchange has to be averaged over all
the fixed points of the orbifold action \cite{U1anomaly},
\begin{equation}
\label{Almwilson} 
A_{lm}^{\alpha \beta} = \frac{i}{|P|} \frac{1}{F} \sum_{k} 
\sum_{f}~C_{k}^{\alpha
\beta}(v) \mbox{Tr}(\gamma_{f}^{\alpha} \lambda^{(0)}_{l})
\mbox{Tr}((\gamma_{f}^{\beta})^{-1} (\lambda^{(0)}_{m})^2) ~, 
\end{equation}
where $F$ is the total number of fixed points, $f$ runs over all
the fixed points of the $k$th twisted sector and $\gamma_{f}$ is the
modified orbifold action at each fixed point due to Wilson line
actions. For example, in a $Z_3$ orientifold model with one Wilson
line, the $27$ fixed points of the $Z_3$ action split into $3$ sets,
each of them feeling different monodromy, $\gamma_{Z_3}$,
$\gamma_{Z_3}\gamma_{W}$ and $\gamma_{Z_3} \gamma_{W}^2$,
respectively. In the present model, the situation is even more
complicated as we will discuss later.

\noindent{\bf (i) Anomaly cancellation of the \boldmath
$Z_3 \times Z_2 \times Z_2$ \unboldmath model without Wilson lines} 

The four U(1)'s from each of the U(6) groups are anomalous. The
mixed triangular anomalies between the four U(1)'s and the SU(6)
and Sp(4) groups from each brane sector can be easily calculated
from the particle spectrum \cite{wilsonlinemodel}
\begin{equation} 
\tilde{A} = \left( \begin{array}{cccccccc}
\;9 & -9 & \;3 & -3 & \;3 & -3 & \;3 & -3 \\[1ex]
\;3 & -3 & \;9 & -9 & \;3 & -3 & \;3 & -3 \\[1ex]
\;3 & -3 & \;3 & -3 & \;9 & -9 & \;3 & -3 \\[1ex]
\;3 & -3 & \;3 & -3 & \;3 & -3 & \;9 & -9\end{array} \right) \;,
\end{equation}
where the row $l$ labels the U(1) from the $D9$ and $D5_i$ brane
sectors and the column $m$ labels the SU(6) and Sp(4) groups. As can
be seen from table \ref{traces_table}, only $Z_3$ twisted moduli
contribute to the anomaly cancellation process and eq.~(\ref{Alm})
indeed cancels the anomalies exactly. For example, $A_{1\,1}$
\begin{equation}
A_{1\,1}=\frac{i}{24} \times (-3\sqrt{3})\times 2 \times 
[2i\sqrt{3}\mbox{Tr}B_{1}^9(-2\mbox{Tr}(B_{1}^{9})^2) ] = -9\;, 
\end{equation}
where we used the standard SU(N) normalization $\mbox{Tr}B_{1}^2=1/2$
for the SU(6) generator, while the U(1) generators are unnormalized.
(We concentrate on the cancellation of mixed anomalies between the
U(1) groups and the non-Abelian symmetries, the U$(1)^3$ anomaly and
the mixed anomalies between U$(1)$ and gravity are cancelled in a
similar way \cite{U1anomaly}.) 

\noindent{\bf (ii) Anomaly cancellation of the \boldmath
$Z_2 \times Z_2 \times Z_3$ \unboldmath model with one Wilson line} 

The field theoretical anomalies between U$(1)$'s and the
non-Abelian groups in this model can be summarized in the following 
matrix
\begin{equation}
\label{Awilson}
\tilde{A} = \left( \begin{array}{cccccccccccc}
\;6 & \;0 & \;0 &  -6 & \;2 & -2 & \;2 & -2 & \;2 & \;0 & \;0 &  -2 \\[1ex]
\;0 & \;1 &  -1 & \;0 & \;1 & -1 & \;1 & -1 & \;0 & \;0 & \;0 & \;0 \\[1ex]
\;3 & \;3 &  -3 &  -3 & \;9 & -9 & \;3 & -3 & \;3 & \;3 &  -3 &  -3 \\[1ex]
\;3 & \;3 &  -3 &  -3 & \;3 & -3 & \;9 & -9 & \;3 & \;3 &  -3 &  -3 \\[1ex]
\;2 & \;0 & \;0 &  -2 & \;2 & -2 & \;2 & -2 & \;6 & \;0 & \;0 &  -6 \\[1ex]
\;0 & \;0 & \;0 & \;0 & \;1 & -1 & \;1 & -1 & \;0 & \;1 &  -1 & \;0
\end{array} \right)\;,
\end{equation}
where the row $l$ denotes U$(1)$ factors from the U$(4)$ and
U$(2)$ of the $D9$ brane sector, the U$(6)$'s of the $D5_{1,2}$ brane
sector, and the U$(4)$ and U$(2)$ of the $D5_3$ brane sector. The
columns count the non-Abelian groups, the order follows table
\ref{matter_table}.

The introduction of the Wilson line implies a more involved  pattern of
anomaly cancellation. In the following we separately address the anomaly
cancellation for different sets of ``anomalous'' U$(1)$'s.

\noindent{\bf (1) The anomalies involving the U(1) \boldmath
$\subset$ \unboldmath U(4)}\\
As shown in table \ref{traces_w_table}, non-zero contributions to
anomaly cancellation, as specified by eq.~(\ref{Almwilson}) come {\it
only} from the $Z_3$ twisted sectors, e.g., in the case of the
mixed anomaly U(1)$\times$SU(4)$^2$, where both of the gauge groups are
from the $99$ sector, eq.~(\ref{Almwilson}) yields
\begin{equation}
A_{1\,1}= \frac{i}{24} \times \frac{1}{3}\times (-3\sqrt{3})\times 2 \times [3 \times 2i\sqrt{3}\mbox{Tr}\tilde{B}_{2}^9(-2\mbox{Tr}(\tilde{B}_{2}^{9})^2)] = -6\;, 
\end{equation} 
where we have used $C_{1}^{99}=-C_{2}^{99}=-3\sqrt{3}$ for the $Z_3$
twist, and $\mbox{Tr}\tilde{B}_2 =4$ for the un\-normalized U(1) factor and
$\mbox{Tr}\tilde{B}_2^2 =1/2$ for the properly normalized SU(4)
generator. Similarly, using the results in table \ref{traces_w_table},
we find that all the entries $\tilde{A}_{1\,2}-\tilde{A}_{1\,4}$ are
exactly cancelled by $A_{lm}$.  However, for the anomalies U$(1)\times
G_i^2$, where $G_i$ comes from the $D5_1$ or $D5_2$ branes, the
situation is different. Since we assume that the $D5_1$ and $D5_2$
branes sit at the origin of the third two-torus (as well as the origin
of the second and first two-tori), they do not feel the action of the
Wilson line, which is acting on the third complex plane. Therefore,
the result from eq.~(\ref{Almwilson}) is simply
\begin{equation}
A_{1\,5}=\frac{i}{24} \times (-\sqrt{3})\times 2 \times 2i\sqrt{3}\mbox{Tr}\tilde{B}_{2}^9(-2\mbox{Tr}(B_{1}^{5_1})^2)=-2\;, 
\end{equation}
where we have $C_{1}^{95_i}= -C_{1}^{95_i}=-\sqrt{3}$ as in the case
without Wilson line, and $B_{1}$ is the generator for the $U(6)$ group
in the $D5_1$ brane sector. Similarly, $\tilde{A}_{1\,6} -
\tilde{A}_{1\,8}$ involving the gauge groups Sp(4) (in $D5_1$ brane
sector), SU(6) and Sp(4) (in $D5_2$ brane sector) are
cancelled. $\tilde{A}_{1\,9}-\tilde{A}_{1\,12}$ involve $G_{i}$ from the
$D5_3$ brane sector, where we again have to take the Wilson line action
into account.  The cancellation works in the same way as in the case
of $\tilde{A}_{1\,1}-\tilde{A}_{1\,4}$. For example,
\begin{equation}
A_{1\,9}= \frac{i}{24} \times \frac{1}{3}\times (-\sqrt{3})\times 2 \times 
[3 \times 2i\sqrt{3}\mbox{Tr}\tilde{B}_{2}^9(-2\mbox{Tr}\tilde({B}_{2}^{5_3})^2)] = -2\;. 
\end{equation}

\noindent{\bf (2) Anomalies involving U(1)
\boldmath$\subset$ \unboldmath U(2)}\\
The relevant traces in table \ref{traces_w_table} show that in the
presence of the background Wilson line, not only the $Z_3$ twisted
sectors, but also the order $6$ twists generated by $Z_3 \times R_3$
contribute to the anomaly cancellation. The latter, however, comes
with different $C_k$ factors. Since the order $6$ twist is generated
by the action diag$(e^{\frac{-2 \pi i}{6}}, e^{\frac{-2 \pi i}{6}},
e^{\frac{2 \pi i}{3}})$, $C_{1}^{99}= -C_{5}^{99}= \sqrt{3}$. Hence,
$A_{2\,2}$ takes the form
\beqa
A_{2\,2} & = & \frac{i}{24} \times \left\{ \frac{1}{3}\times 
(-3\sqrt{3})\times 2 \times 
[2i\sqrt{3}\mbox{Tr}\tilde{B}_{1}^9(-2\mbox{Tr}(\tilde{B}_{1}^{9})^2) \right.
\NO\\[1ex]
& & + 2 \times 2i\sqrt{3}\mbox{Tr}\tilde{B}_{1}^9(-\mbox{Tr}(\tilde{B}_{1}^{9})^2)] 
\NO\\[1ex]
& & \left.+ \frac{1}{3}\times (\sqrt{3})\times 2 \times 
    [2 \times (-3)\mbox{Tr}\tilde{B}_{1}^9(i \sqrt{3}
    \mbox{Tr}(\tilde{B}_{1}^{9})^2)] \right\}\NO\\[1ex]
& = & -1\;.
\eeqa
One sees that among the various contributions to $A_{2\,1}$, in the $Z_3$
twisted sectors the contribution from fixed points which do not feel
the Wilson line cancel those from fixed points which do feel the
Wilson line; while in $Z_6$ twisted sectors, the contributions from
fixed points that feel the Wilson line action cancel between each
other. As a net result, $A_{2\,1}=0$. $A_{2\,3}$ and $A_{2\,4}$ cancel
their counterparts from $\tilde{A}$ as well. In cancelling
$\tilde{A}_{2\,5}-\tilde{A}_{2\,8}$ in which the non-Abelian group
arises from the $5_1$ or $5_2$ sector, the Wilson line action does not
have any effect, thus $\tilde{A}_{2\,5}$ is cancelled by
\begin{equation}
A_{2~5} =\frac{i}{24} \times (-\sqrt{3}) \times 2 \times 2i\sqrt{3} 
\mbox{Tr} \tilde{B}_{1}^{9} (-2 \mbox{Tr}(B_{1}^{5_1})^{2}) = -1\;. 
\end{equation}
When $G_i$ comes from the $D5_3$ brane sector, the effects from the
$Z_6$ twists again have to be taken into account. The net result is
that the contribution from the $Z_3$ twists cancels those from the 
$Z_6$ twists in such a way that $A_{2~9}-A_{2~12}=0$. 

\noindent{\bf (3) Anomalies involving U(1) \boldmath$\subset$
\unboldmath U(6)}\\
Since the U$(6)$ group arises from the $D5_1$ or $D5_2$ brane sectors,
where the Wilson line acts trivially, the cancellation of anomalies of 
this type is the same as in the original model without Wilson line.     

We have also checked explicitly all the U$(1)^3$ anomalies in the
model, and confirmed that they are exactly cancelled by the
generalized GS mechanism.     

\section{Fayet-Iliopoulos terms and gauge coupling corrections}

\subsection{Fayet-Iliopoulos terms}
The supersymmetric completion of the term (\ref{1thCS}) generates the
Fayet-Iliopoulos (FI) terms for the associated anomalous
U(1)'s.~\footnote{Note that there could be sigma-model anomaly
corrections to the FI-terms
\cite{IbaneztypeI}, proportional to the untwisted sector moduli. (See 
also Ref.~\cite{Klein}.)} In the case of $Z_2\times Z_2 \times Z_3$
orientifold, the FI term of the U$(1)_A$ in the $D9$ brane sector
takes the form
\begin{equation}
\xi_{FI}^{9} \sim \int d^4 x ~\mbox{Tr}(\gamma_{Z_3} \lambda^{(0)}) 
\sum_{i,j,k=1}^{3} \phi_{i,j,k} \;,
\end{equation} 
where $\phi_{i,j,k}$ are the NS-NS sector moduli arising from the
$Z_3$ twisted sectors, since, as we argued earlier only $Z_3$ twisted
sectors contribute to the anomaly cancellation process. (Notice that
we are summing over only half of the twisted sectors.) The FI terms of
the anomalous U$(1)$ from the $5_1$ branes involve the NS-NS moduli
$\phi_{i,1,1}$, those from $5_2$ branes involve $\phi_{1,i,1}$, etc..
In the following, we discuss the FI terms when a background Wilson line
is added.

\noindent {\bf (a) The FI term of the U(1)'s from 
\boldmath$D9$ \unboldmath branes}\\
For the anomalous U(1) $\subset$ U(4), 
\beqa
\xi_{FI,1}^{9} & \sim & \sum_{i=1}^{3} \sum_{j=1}^{3} \;\left[ \mbox{Tr}(\gamma_{Z3} 
\lambda^{(0)}) \phi_{i,j,1} +  \mbox{Tr}(\gamma_{Z3}\gamma_{W} 
\lambda^{(0)})\phi_{i,j,2} + \mbox{Tr}(\gamma_{Z3}\gamma_{W}^2 
\lambda^{(0)})\phi_{i,j,3}\right] \NO\\[1ex]
& = & (2i\sqrt{3}) \mbox{Tr}\tilde{B}_{2}^{9} 
\sum_{i,j,k=1}^{3} (\phi_{i,j,k})\;, 
\eeqa
where $\phi_{i,j,k}$ are moduli fields from $Z_3$ twisted sectors and
the overall normalization factor has been suppressed. Note that,
although the blowing-up modes $\phi_{i,j,k}$ form a convenient basis,
they are not physical moduli of $Z_3$. As discussed
earlier, only a specific combination of nine blowing-up modes is
physical.

In the case of U(1) $\subset$ U(2), the order $6$ twists also contribute, 
\beqa
\xi_{FI,2}^{9} & \sim & \sum_{i=1}^{3} \sum_{j=1}^{3} [ \mbox{Tr}(\gamma_{Z3}
\lambda^{(0)}) \phi_{i,j,1} +  \mbox{Tr}(\gamma_{Z3}\gamma_{W}
\lambda^{(0)})\phi_{i,j,2} + \mbox{Tr}(\gamma_{Z3}\gamma_{W}^2
\lambda^{(0)})\phi_{i,j,3}] \NO\\[1ex]
&   &  + [ \mbox{Tr}(\gamma_{Z3} \gamma_{R3} \lambda^{(0)}) \rho_{1,1,1} +
\mbox{Tr}(\gamma_{Z3} \gamma_{R3} \gamma_{W} \lambda^{(0)})\rho_{1,1,2} +
\mbox{Tr}(\gamma_{Z3} \gamma_{R3}\gamma_{W}^2 \lambda^{(0)})\rho_{1,1,3}]
\NO\\[1ex]
& = & (i\sqrt{3}) \mbox{Tr}\tilde{B}_{1}^{9} \sum_{i=1}^{3} \sum_{j=1}^{3} (2\phi_{i,j,1} - \phi_{i,j,2} - \phi_{i,j,3}) + (-3i) \mbox{Tr}\tilde{B}_{1}^{9} ( \rho_{1,1,2}- \rho_{1,1,3})\;.
\eeqa
 
\noindent{\bf (b) The FI terms of the U(1)'s from 
\boldmath$D5_3$ \unboldmath branes}\\
 Since the $5_3$ sector is subject to the Wilson line action, the
basic forms of the FI terms are the same as in the $D9$ brane sector. 
However, since $5_3$ branes sit at the origin of the 1st and 2nd
complex tori, only those moduli with $i=j=1$ are involved. Hence, 
\begin{equation} 
\xi_{FI,1}^{5_3}  \sim (2i\sqrt{3}) \mbox{Tr}\tilde{B}_{2}^{5_3} (\phi_{1,1,1} + \phi_{1,1,2} + \phi_{1,1,3})\;, 
\end{equation}
is the FI term for the U(1) $\subset$ U(4). On the other hand, 
\begin{equation}
\xi_{FI,2}^{5_3}  \sim (i\sqrt{3}) \mbox{Tr}\tilde{B}_{1}^{5_3} (2\phi_{1,1,1} - \phi_{1,1,2} - \phi_{1,1,3}) + (-3i) \mbox{Tr}\tilde{B}_{1}^{5_3} ( \rho_{1,1,2}- \rho_{1,1,3})\;,  
\end{equation}
gives the FI term for the U(1) $\subset$ U(2). 

\noindent{\bf (c) The FI terms of the U(1)'s that arise from 
\boldmath$D5_1$ and $D5_2$ \unboldmath branes}\\
As suggested by the anomaly cancellations, only moduli associated to 
the $Z_3$ twists are relevant. Hence, the FI term for the U(1) in
the $5_1$ brane sector is 
\begin{equation} 
\xi_{FI}^{5_1}  \sim (2i\sqrt{3}) \mbox{Tr}\tilde{B}_{1}^{5_1} (\phi_{1,1,1} + \phi_{2,1,1} + \phi_{3,1,1})\;, 
\end{equation}
and for the U(1) from $D5_2$ branes, 
\begin{equation} 
\xi_{FI}^{5_2}  \sim (2i\sqrt{3}) \mbox{Tr}\tilde{B}_{1}^{5_2} (\phi_{1,1,1} + \phi_{1,2,1} + \phi_{1,3,1})\;. 
\end{equation}

\subsection{Gauge coupling corrections}

The gauge kinetic functions receive corrections from the twisted
sector blowing-up modes, in a mechanism related to the Chern-Simons
term \cite{IbaneztypeI,CELW,dudas,nilles}; they have their origin in
the second-order corrections, and the proposed form of the corrections
\cite{IbaneztypeI} was confirmed by recent explicit string
calculations \cite{dudas}. The holomorphic gauge coupling function is
of the form:
\begin{equation}
f= S + \delta f(R),
\end{equation}
where $S$ is the (untwisted sector) dilaton for $D9$ brane sector or 
the untwisted toroidal modulus $T_i$ for $D5_i$ brane sectors. For a $Z_N$ 
orbifold, the correction can be expressed as 
\begin{equation}
\label{deltaf}
\delta f(R) \sim \sum_{k} \mbox{Tr}(\gamma_{\theta^k} \lambda^{(0)2}) \sum_{l,m,n}\phi_{l,m,n}, 
\end{equation} 
where the summation is over half of the twisted sectors, and
$\phi_{l,m,n}$ are the NS-NS twisted sector moduli associated with the
particular twist. (Again an overall normalization factor is
suppressed.) In the case of the $Z_{2}\times Z_2 \times Z_3$ model,
$\delta f$ receives non-trivial contributions from $Z_3$ twists only
(cf. table \ref{traces_table}), such that
\begin{equation}
\delta f_1^9 \sim -2 \mbox{Tr} (B_{2}^{9})^2 
\sum_{i,j,k=1}^{3}(\phi_{i,j,k})\;,
\end{equation}
and 
\begin{equation}
\delta f_2^9 \sim 4 \mbox{Tr} ((B_2^9)^2+S_2^9 S_1^9) \sum_{i,j,k}^{3}(\phi_{i,j,k})\;, 
\end{equation}
for the SU(6) and Sp(4) groups in the $D9$ brane sector.  The
corrections take a similar form for gauge groups arising from the
$D5_{i}$ brane sector, except that the summations over $i,j,k$ are
replaced by fixed values $j=k=1$, $k=i=1$ and $i=j=1$, respectively.

With the Wilson line, in the $D9$ brane sector, table \ref{traces_w_table} 
shows that the
gauge coupling of SU(4) receives non-zero corrections, as specified
by eq.~(\ref{deltaf}), only from $Z_3$ twists. It takes the form 
\begin{equation}
\delta f_1^9 \sim -2 \mbox{Tr} (\tilde{B}_{2}^{9})^2 
\sum_{i,j,k=1}^{3}(\phi_{i,j,k})\;,
\end{equation}
while the gauge coupling correction for SU(2) ($\subset$ U(2)),
gets contributions from order three, {\it and} order six twisted 
sectors. It is of the form  
\begin{equation}
\delta f_2^9 \sim - \mbox{Tr} (\tilde{B}_{1}^{9})^2 
\sum_{i,j}^{3}(2\phi_{i,j,1}-\phi_{i,j,2}-\phi_{i,j,3} ) +  
 \sqrt{3} \mbox{Tr} (\tilde{B}_{1}^{9})^2 (\rho_{1,1,2} - \rho_{1,1,3})\;.    
\end{equation}
Notice that from table \ref{traces_w_table}, the order $2$ twist $R_3$
appears to have a contribution to eq.~(\ref{deltaf}) with the relevant
traces being non-zero. However, since $R_3$ acts trivially on the third
complex torus, the Wilson line action does not differentiate the moduli of
the $R_3$ twist. Hence, the total contribution from the $R_3$ sector 
to $\delta f$ vanishes after summing over $\gamma_{W}^k$.    

Similarly, for the first SU(2)$\equiv$Sp(2) group in the $D9$ brane
sector one obtains  
\begin{equation}
\delta f_3^9 \sim 2( (W_1^9)^2+W_2^9 W_3^9) \sum_{i,j}^{3}(2\phi_{i,j,1}
-\phi_{i,j,2}-\phi_{i,j,3} ) - 2  \sqrt{3} ((W_1^9)^2+W_2^9 W_3^9) 
(\rho_{1,1,2} - \rho_{1,1,3})\;.    
\end{equation}
And the second $SU(2)$ receives the following correction to its
gauge kinetic function:
\begin{equation}
\delta f_4^9 \sim 4((Z_1^9)^2+Z_2^9 Z_3^9) \sum_{i,j,k}^{3}(\phi_{i,j,k})\;. 
\end{equation}

The corrections to the gauge couplings in the $D5_3$ brane sector are very
similar to those in the $D9$ brane sector, except that the summation
over $i,j$ is changed to a fixed value $i=j=1$. 

In the $D5_1$ sector, the correction to the gauge coupling of SU(6)
is simply
\begin{equation}
\delta f_1^{5_1} \sim -2 \mbox{Tr}(B_{1}^{5_1})^2 \sum_{i=1}^3 (\phi_{i,1,1}), 
\end{equation}
and that of Sp(4) is 
\begin{equation}
\delta f_2^{5_1} \sim 4 \mbox{Tr}((B_{2}^{5_1})^2 + S_{2}^{5_1} S_{1}^{5_1}) \sum_{i=1}^3 (\phi_{i,1,1}). 
\end{equation} 
The corrections to the gauge functions in the $D5_2$ brane sector are 
similar to those in the $D5_1$ sector, except that the moduli fields 
which are involved are $\phi_{1,i,1}$.  
  
\section{Conclusions}
We have presented an explicit construction of a $N=1$ supersymmetric,
three family, Type IIB $Z_3\times Z_2\times Z_2$ orientifold with a
Wilson line that does not commute with the orbifold group.  One of the
motivations for this construction was to find a model with a gauge
group structure that is close to that of the standard model and thus
provide an example of a model with potentially quasi-realistic
features.  Unfortunately, the only examples that are free of
non-Abelian anomalies and are consistent with tadpole cancellation,
while still containing the SM gauge group as a subgroup, have a gauge
group structure of the type U(4)$\times$U(2)$\times$SU(2)$
\times$SU(2)($\times$U(4)$\times$U(2)$\times$SU(2)$\times$SU(2)$
\times$(U(6)$\times$Sp(4))$^{2}$), and thus its ``observable'' sector 
gauge structure is still much larger than that of the SM. (While our
study builds on an earlier work \cite{wilsonlinemodel}, we found a
different gauge group structure and massless spectrum).

The gauge group U(4)$\times$U(2)$\times$SU(2)$\times$SU(2) can be
easily understood in the T-dual picture, in which the Wilson line
action on the $D5_3$ branes is dual to splitting the $32$ $D5_3$
branes and placing them at various fixed points, while respecting the
orbifold symmetries. In the T-dual picture, a set of $20$ $D5_3$
branes is placed at the origin, which gives rise to the gauge group
U(4)$\times$Sp(2) when the $Z_{3} \times Z_{2} \times Z_{2}$
orientifold actions are imposed. A set of $6$ $D5_3$ branes is placed
at one of the two fixed points of $Z_3$ in the third complex two-torus
which are away from the origin. Due to the two $Z_2$ actions, another
set of $6$ $D5_3$ branes needs to be placed at the other fixed point
as the mirror image of the first set of $6$. These two sets of $6$
$D5_3$ branes yield only one set of physical states and give rise to
the gauge group U(2)$\times$Sp(2) under the additional $Z_3$ orbifold
action.

In addition to obtaining the gauge group structure, the massless
spectrum and the trilinear superpotential, we carried out the explicit
calculation of the Abelian anomaly cancellations, employing a
generalization of the Green-Schwarz mechanism (as proposed in
ref.~\cite{U1anomaly} and confirmed in ref.~\cite{dudas}).  In
particular we emphasized subtleties associated with ``anomalous''
U(1)'s of U(4) and U(2) group factors, which are due to the
non-trivial role non-Abelian Wilson lines are playing in the anomaly
cancellation.  In addition, we also calculated the FI-terms and the
gauge coupling corrections due to the blowing-up modes, thus setting a
stage for further investigations of the physics implications of this
type of models.

\enlargethispage*{1cm}
The techniques we have employed in order to obtain a specific anomaly
free three family model can be easily applied to the study of a
broader class of Type IIB orientifolds, i.e.\ models with other
orbifold groups, and with a more general form of the non-Abelian
Wilson lines.  In addition, a systematic formulation of the
consistency constraints for this class of string solutions is clearly
an important question, and deserves further study.

\acknowledgments

We would like to thank Angel Uranga for a collaboration on parts of
the work presented in the paper and for his many insightful
communications and suggestions. We also benefited from discussions
with L.~Everett and P.~Langacker.  This work was supported in part by
U.S.\ Department of Energy Grant No.~DOE-EY-76-02-3071 (M.C.\ and
M.P.), Grant No.~DE-AC02-76CH03000 (J.W.), in part by the University
of Pennsylvania Research Foundation award (M.C.), and in part by the
Feodor Lynen Program of the Alexander von Humboldt Foundation (M.P.).

\clearpage

\begin{appendix}

\section{The matter fields \protect\label{MATTER}}
\noindent
{\bf 99-sector:}\\
Like for the gauge fields, the Chan-Paton matrices of the matter
states have to be invariant under the action of the orientifold
group. Since the world-sheet states of the matter fields are not
invariant under the orientifold action, eqs.~(\ref{ori1}) and 
(\ref{ori2}) now imply that
\beq
  \lambda^{(i)}=-\gamma_{\Omega,9}{\lambda^{(i)}}^{\tr}\gamma_{\Omega,9}^{-1} 
  \qquad\mbox{and}\qquad
  \lambda^{(i)}=\mbox{e}^{2i\pi v_i}\gamma_{g,9}\lambda^{(i)}\gamma_{g,9}^{-1}\;.
\eeq
{\bf\boldmath$5_i5_i$-sector\unboldmath}\\
Similarly, if open strings begin and end on $D5_i$ branes, the 
Chan-Paton matrices have to obey the following constraints:
\beqa
  \lambda^{(i)}=-\gamma_{\Omega,5}{\lambda^{(i)}}^{\tr}\gamma_{\Omega,5}^{-1}\;, &\quad\qquad&
  \lambda^{(i)}=\mbox{e}^{2i\pi v_i}\gamma_{g,5_i}\lambda^{(i)}\gamma_{g,5_i}^{-1}\;,\\
  \lambda^{(j)}=+\gamma_{\Omega,5}{\lambda^{(j)}}^{\tr}\gamma_{\Omega,5}^{-1}\;, &\quad\qquad&
  \lambda^{(j)}=\mbox{e}^{2i\pi v_j}\gamma_{g,5_i}\lambda^{(j)}\gamma_{g,5_i}^{-1}\;,
  \quad\mbox{for }i\ne j\;.
\eeqa
The sign change in the world-sheet parity projection in the last line 
stems from the DD boundary conditions in the $j\ne i$ directions
transverse to the $D5_i$ branes.\\
{\bf\boldmath$95_i$-sectors\unboldmath}: \\
Further, one can have mixed states where the open strings begin and
end on different branes. In the $95_i$ case, the coordinates obey
mixed DN boundary conditions, i.e.\ they have half-integer
modings. The states can be written as $\ket{s_j,s_k,ab}\lambda_{ab}$,
$j,k\ne i$, with helicities $s_j,s_k=\pm{1\over2}$, and the Chan-Paton
index $a$ ($b$) lies on a $D5$ brane ($D9$ brane).  Due to the GSO
projection one has $s_j=s_k$.  The orientifold projections now imply
\beq
  \lambda=\mbox{e}^{2\pi i(v_js_j+v_ks_k)}\gamma_{g,5_i}\lambda\gamma_{g,9}^{-1}\;.
\eeq 
The world-sheet parity operation $\Omega$ only relates $95_i$ and $5_i9$
sectors and does not yield any additional constraints.\\
{\bf\boldmath$5_j5_i$-sectors\unboldmath}:\\
Finally, the matter states in mixed $5_j5_i$-sectors are determined
by: 
\beq
  \lambda=\mbox{e}^{2\pi i(v_is_i+v_js_j)}\gamma_{g,5_i}\lambda\gamma_{g,5_j}^{-1}\;.
\eeq 

As an example, consider the $Z_2\times Z_2\times Z_3$ orientifold
discussed in the main text. In the 99-sector the matter Chan-Paton 
matrices have to satisfy
\beqa
 \lambda^{(1)}=\mbox{e}^{2i\pi/3}\gamma_{Z_3,9}\lambda^{(1)}\gamma_{Z_3,9}^{-1}\;,&\quad&
 \lambda^{(1)}=+\gamma_{R_1,9}\lambda^{(1)}\gamma_{R_1,9}^{-1}\;,\quad
 \lambda^{(1)}=-\gamma_{R_2,9}\lambda^{(1)}\gamma_{R_2,9}^{-1}\;,\\[1ex]
 \lambda^{(2)}=\mbox{e}^{2i\pi/3}\gamma_{Z_3,9}\lambda^{(2)}\gamma_{Z_3,9}^{-1}\;,&\quad&
 \lambda^{(2)}=-\gamma_{R_1,9}\lambda^{(2)}\gamma_{R_1,9}^{-1}\;,\quad
 \lambda^{(2)}=+\gamma_{R_2,9}\lambda^{(2)}\gamma_{R_2,9}^{-1}\;,\\[1ex]
 \lambda^{(3)}=\mbox{e}^{2i\pi/3}\gamma_{Z_3,9}\lambda^{(3)}\gamma_{Z_3,9}^{-1}\;,&\quad&
 \lambda^{(3)}=-\gamma_{R_1,9}\lambda^{(3)}\gamma_{R_1,9}^{-1}\;,\quad
 \lambda^{(3)}=-\gamma_{R_2,9}\lambda^{(3)}\gamma_{R_2,9}^{-1}\;.
\eeqa
Imposing these constraints yields, e.g.\ for $\lambda^{(1)}$
\beq
\lambda^{(1)}=\left(\begin{array}{cccc}
   0           &  0  &\sigma_3\otimes A& 0 \\[1ex]
\sigma_1\otimes M_1&    0     &   0      &   0 \\[1ex]
   0           &-\sigma_3\otimes M_2^{\tr}&   0      &-\sigma_1\otimes M_1^{\tr}\\[1ex]
\sigma_3\otimes M_2&  0    &    0     &   0     
  \end{array}\right)\;,
  \label{l1_1}
\eeq
where $A$ is an antisymmetric $6\times6$ matrix and $M_1$ and $M_2$
are general $2\times6$ matrices. Under the gauge group
U(6)$\times$Sp(4), $A$ transforms as $(15,1)_{+2}$, while $M_1$ and
$M_2$ form a $(\Bar{6},4)_{-1}$, where the subscripts $+2$ and $-1$
denote the U(1) charges of the states.  $\lambda^{(2)}$ and $\lambda^{(3)}$ have
a similar structure, i.e.\ we get three copies of each of these states
\cite{wilsonlinemodel}.

Introducing a Wilson line $\gamma_W$ yields an additional constraint
\beq
  \lambda^{(i)}=\gamma_W \lambda^{(i)} \gamma_W^{-1}\;,
\eeq
for those states that feel the Wilson line action. In the case of one
Wilson line along the third two-torus, considered in the main text,
this would only affect states in the 99-, $5_35_3$- and
$95_3$-sectors, since we assumed that all the $D5$ branes are located at
the fixed point at the origin, where the Wilson line is not active.

For example, in the case of $\lambda^{(1)}$ from the 99-sector the matrices 
$A$, $M_1$ and $M_2$ from eq.~(\ref{l1_1}) get reduced to
\beq
  A=\left(\begin{array}{cc} 
          0 & 0 \\[1ex] 0 & \tilde{A} 
    \end{array}\right)\;,
  \qquad
  M_1=\left(\begin{array}{cc} 
          0 & 0 \\[1ex] 0 & \tilde{M}_1 
    \end{array}\right)\;,
  \qquad
  M_2=\left(\begin{array}{cc} 
          0 & 0 \\[1ex] 0 & \tilde{M}_2
    \end{array}\right)\;,
  \label{l1_2}
\eeq
where $\tilde{A}$ is an antisymmetric $4\times4$ matrix and 
$\tilde{M}_1$ and $\tilde{M}_2$ are general 4-dimensional
line-vectors. Under the gauge group
U(4)$\times$U(2)$\times$SU(2)$\times$SU(2), $\tilde{A}$ transforms as
$(6,1,1,1)(+2,0)$ and $(\tilde{M}_1,\tilde{M}_2)$ form a
$(\bar{4},1,1,2)(-1,0)$, where the second set of brackets contains the 
U(1) charges of the fields.

A novel feature of this Wilson line action is that now the
three matter Chan-Paton matrices in the 99- and $5_35_3$-sectors 
yield different matter states. Consider $\lambda^{(3)}$ from the
99-sector. In the absence of Wilson lines it will give the same matter
states $A$, $M_1$ and $M_2$ as $\lambda^{(1)}$ (cf.~eq.~(\ref{l1_1})). 
However, Wilson line action on $\lambda^{(3)}$ projects out fewer states
than in the case of $\lambda^{(1)}$ and $\lambda^{(2)}$ and the surviving fields
read
\beq
  A=\left(\begin{array}{cc} 
          \tilde{A}_2 & 0 \\[1ex] 0 & \tilde{A} 
    \end{array}\right)\;,
  \qquad
  M_1=\left(\begin{array}{cc} 
          \tilde{M}_3 & 0 \\[1ex] 0 & \tilde{M}_1 
    \end{array}\right)\;,
  \qquad
  M_2=\left(\begin{array}{cc} 
          \tilde{M}_4 & 0 \\[1ex] 0 & \tilde{M}_2
    \end{array}\right)\;,
  \label{l3_2}
\eeq
where, in addition to the states $(6,1,1,1)(+2,0)$ and
$(\bar{4},1,1,2)(-1,0)$ discussed above, we have an antisymmetric
$2\times2$ matrix $\tilde{A}_2$, corresponding to a $(1,1,1,1)(0,+2)$,
and two-dimensional line vectors $\tilde{M}_3$ and $\tilde{M}_4$ which 
form a $(1,2,2,1)(0,-1)$.

\section{Blowing-up modes from the $Z_2$ twisted sectors in 
$Z_3\times Z_2 \times Z_2$ orbifolds} 

The physical moduli arising from each of the three $Z_2$ twisted
sectors have a similar form, we present the explicit result for the
$R_3$ twist only. Let us denote the twist fields associated with the
four ($i=1,2,3,4$) fixed points of the first two two-tori ($J=1,2$)
under $R_3$ action as $\Sigma_{i}^{J}$. The blowing-up modes of the
total $16$ fixed points are represented by the fields
$\omega_{i,j}=\Sigma_{i}^{1}\Sigma_{i}^{2}$. However, only the
following six combinations of $\omega_{i,j}$ are invariant under the
$Z_3$ rotation:
\beqa
\omega_{1,1} \;,&& \\
\omega_{\{\{2,3,4\}\}, \{\{2,3,4\}\}^*}&\equiv&
{1\over {3}}(\Sigma_{2}^{1}+e^{{2\pi i}\over
3}\Sigma_{3}^1+e^{{4\pi i}\over 3}\Sigma_{4}^{1})(\Sigma_{2}^{2}+e^{{4\pi
i}\over 3}\Sigma_{3}^{2}+e^{{2\pi i}\over 3}\Sigma_{4}^{2})\;,
\\
\omega_{\{\{2,3,4\}\}^*, \{\{2,3,4\}\}}&\equiv& {1\over {3}}(
\Sigma_{2}^{1}+e^{{4\pi i}\over
3}\Sigma_{3}^{1}+e^{{2\pi i}\over 3}\Sigma_{4}^{1})(\Sigma_{2}^{2}+e^{{2\pi
i}\over 3}\Sigma_{3}^{2}+e^{{4\pi i}\over 3}\Sigma_{4}^{2})\;,
\\
\omega_{\{2,3,4\}, 1}&\equiv&{1\over {\sqrt 3}}(\Sigma_{2}^{1}+\Sigma_{3}^{1}+
\Sigma_{4}^{1}) \Sigma_{1}^{2}\;,
\\
\omega_{1,\{2,3,4\}}&\equiv&{1\over {\sqrt 3}}\Sigma_{1}^{1}
(\Sigma_{2}^{2}+\Sigma_{3}^{2}+\Sigma_{4}^{2})\;,
\\
\omega_{\{2,3,4\}, \{2,3,4\}}&\equiv& {1\over {3}}(\Sigma_{2}^{1}+
\Sigma_{3}^{1}+\Sigma_{4}^{1})(\Sigma_{2}^{2}+\Sigma_{3}^{2}+\Sigma_{4}^{2})\;.
\eeqa

\end{appendix}

\clearpage

\def\B#1#2#3{\/ {\bf B#1} (19#2) #3}
\def\NPB#1#2#3{{\it Nucl.\ Phys.}\/ {\bf B#1} (19#2) #3}
\def\PLB#1#2#3{{\it Phys.\ Lett.}\/ {\bf B#1} (19#2) #3}
\def\PRD#1#2#3{{\it Phys.\ Rev.}\/ {\bf D#1} (19#2) #3}
\def\PRL#1#2#3{{\it Phys.\ Rev.\ Lett.}\/ {\bf #1} (19#2) #3}
\def\PRT#1#2#3{{\it Phys.\ Rep.}\/ {\bf#1} (19#2) #3}
\def\MODA#1#2#3{{\it Mod.\ Phys.\ Lett.}\/ {\bf A#1} (19#2) #3}
\def\IJMP#1#2#3{{\it Int.\ J.\ Mod.\ Phys.}\/ {\bf A#1} (19#2) #3}
\def\nuvc#1#2#3{{\it Nuovo Cimento}\/ {\bf #1A} (#2) #3}
\def\RPP#1#2#3{{\it Rept.\ Prog.\ Phys.}\/ {\bf #1} (19#2) #3}
\def\etal{{\it et al\/}}


\end{document}